\documentclass[12pt]{article}
\usepackage{indentfirst}
\usepackage{graphicx}
\usepackage{placeins}
\usepackage{slashed}
\usepackage{gensymb}
\usepackage{amsmath}
\usepackage{amssymb}
\usepackage{feynmp-auto}
\usepackage{multirow}
\usepackage[hidelinks]{hyperref}
\usepackage{cite}

\newcommand\pubnumber{}
\newcommand\pubdate{\today}

\def\napoli{Physikalisches Institut and Bethe Center for Theoretical Physics,\\
Bonn University, 53115 Bonn, Germany}
\def\support{\footnote{zhongyi@th.physik.uni-bonn.de}}

\def\Title#1{\begin{center} {\Large #1 } \end{center}}
\def\Author#1{\begin{center}{ \sc #1} \end{center}}
\def\Address#1{\begin{center}{ \it #1} \end{center}}

\newcommand\pubblock{\rightline{\begin{tabular}{l} \pubnumber\\
         \pubdate  \end{tabular}}}
\newenvironment{Abstract}{\begin{quotation}  }{\end{quotation}}

\def\Acknowledgements{\bigskip  \bigskip \begin{center} \begin{large}
             \bf ACKNOWLEDGEMENTS \end{large}\end{center}}

\begin{document}

\def\gsim{\:\raisebox{-0.5ex}{$\stackrel{\textstyle>}{\sim}$}\:}
\def\lsim{\:\raisebox{-0.5ex}{$\stackrel{\textstyle<}{\sim}$}\:}

\begin{titlepage}
\pubblock

\vfill
\Title{LHC Constraints on a Mediator Coupled to Heavy Quarks}
\Author{Manuel Drees\footnote{drees@th.physik.uni-bonn.de},
Zhongyi Zhang\support}
\Address{\napoli}

\begin{Abstract} 
  We apply LHC data to constrain a simplified extension of the
  Standard Model containing a new spin$-1$ mediator $R$, which does
  not couple to first generation quarks, and a spinor dark matter
  particle $\chi$. We recast ATLAS and CMS searches for final states
  containing one or more jet(s) + $\slashed{E}_T$, with or without $b$
  tags, as well as searches for di--jet resonances with $b$ or $t$
  tagging. We find that LHC constraints on the axial vector couplings
  of the mediator are always stronger than the unitarity bound, which
  scales like $m_R/m_t$. If $R$ has a sizable invisible branching
  ratio, the strongest LHC bound on both vector couplings and axial
  vector coupling comes from a di--jet + $\slashed{E}_T$ search with
  or without double $b$ tag. These bounds are quite strong for
  $m_R < 1$ TeV, even though we have switched off all couplings to
  valence quarks. Searches for a di--jet resonance with double $b$ tag
  lead to comparable bounds with the previous results even if
  $R \rightarrow \chi \bar \chi$ decays are allowed; these are the
  only sensitive LHC searches if the invisible branching ratio of $R$
  is very small or zero.
\end{Abstract}
\vfill

\end{titlepage}
\def\thefootnote{\fnsymbol{footnote}}
\setcounter{footnote}{0}

\section{Introduction}
\label{sec:intro}

Simplified models of particle dark matter often need a mediator
coupling the dark matter particle $\chi$ to some particles in the
Standard Model (SM). Models where the mediator couples to both quarks
and leptons are strongly constrained by LHC searches for
$\ell^+ \ell^-$ resonances, where $\ell$ stands for a charged lepton
\cite{Khachatryan:2016qkc, Aaboud:2017buh, Aaboud:2017sjh,
  Sirunyan:2018exx}.  This motivates the investigation of
``leptophobic'' models, where the mediator does not couple to leptons.
In case of a spin$-1$ mediator $R$, universal couplings to all quarks
are often assumed. If $R$ has a sizable branching ratio into invisible
final states, which is generally true if $m_R > 2 m_\chi$, the allowed
vector and axial vector couplings are then strongly constrained by
mono--jet searches \cite{Sirunyan:2017hci, Aaboud:2017phn} unless
$m_R$ is well above $1$ TeV. For mediator mass between $1$ and $2.5$
TeV, searches for di--jet resonances \cite{Aaboud:2017yvp,
  Sirunyan:2016iap} perform even better. Additionally, the constraints
from spin--dependent and spin--independent interactions in direct
detection experiments imposes strong constraints on couplings to first
generation quarks \cite{Frandsen:2012rk}; these bounds scale like
$m_R$.

In our previous study \cite{Drees:2018gvr}, which applies LEP data to
probe the low $m_R$ region, we therefore switched off all couplings to
first generation quarks and axial vector couplings to second
generation quarks in order to avoid an excess in direct Dark Matter
detection experiments. At tree level, axial vector couplings lead to
spin--dependent contributions to the scattering cross section, which
also receive a sizable contribution from strange quarks, whereas
vector couplings lead to spin--independent contributions which only
probe $u$ and $d$ quarks in the nucleon \cite{Ellis:1991ef}. In our
model the scattering on nuclei can therefore only proceed via loop
diagrams, and should thus be strongly suppressed.\footnote{For purely
  vectorial interaction the effective $Rgg$ vertex should vanish
  according to Furry's theorem; one will have to add a second $R$
  exchange or a third external gluon in order to obtain a
  non--vanishing contribution. In case of axial vector interaction the
  effective $Rgg$ vertex seems to lead to a velocity--suppressed
  contribution to the cross section, so the dominant contribution
  probably again comes from yet higher orders.} Moreover, the
non--zero couplings to other quarks are still available to generate a
sizable annihilation rate to explain the observed dark matter relic
density through thermal freeze--out. By switching off couplings to
first generation quarks, and hence to all valence quarks, we greatly
reduce the cross sections for $pp$ scattering processes with an $R$
boson in the intermediate or final state. The published bounds from
the LHC experiments, which assume equal couplings of $R$ to all
quarks, are therefore no longer valid.

The goal of this article is to estimate the LHC constraints on this
model. We showed in ref.\cite{Drees:2018gvr} that LEP data impose
strong constraints only for $m_R < 10$ GeV, and become entirely
insensitive for $m_R > 70$ GeV. Here we therefore focus on scenarios
with $m_R \geq 10$ GeV. The relevant searches we exploit are similar
to those that constrain scenarios with flavor--universal couplings of
$R$: mono--jet + $\slashed{E}_T$ searches, di--jet + $\slashed{E}_T$
searches and di--jet resonance searches. By switching off couplings to
light quarks, we increase the branching ratio for
$R \rightarrow b \bar b$ or $t \bar t$ decays. Since in background
events most jets originate from light quarks or gluons, $b$ or $t$
tagging can increase the signal to background ratio even for
flavor--universal couplings of $R$, and should be even more helpful in
our case.

The reminder of this article is organized as follows. In
Sec.~\ref{sec:2}, we briefly describe the Lagrangian of the simplified
model containing a leptophobic mediator, which does not couple to
first generation quarks. The application to the relevant LHC data is
discussed in Sec.~3. The LEP result and the tightest unitarity
condition from top quark are compared to the LHC exclusion limits we
estimate. Finally, Sec.~\ref{sec:4} contains our summary and
conclusions.

\section{The Simplified Model}
\label{sec:2}

\subsection{Lagrangian and Free Parameters}
\label{subsec:2.1}

A spinor dark sector particle (DSP) and a new spin$-1$ mediator connecting
DSP to SM particles are introduced to extend the SM. Therefore the
total Lagrangian is given by:
\begin{equation} \label{L_tot}
\mathcal{L} = \mathcal{L}_{\rm SM} + \mathcal{L}_{\rm DSP}
+ \mathcal{L}_R + \mathcal{L}_I\,.
\end{equation}
Since we use MadGraph \cite{alwall2011madgraph} to
generate the Monte Carlo Events, the kinetic terms in the Lagrangian follow
the default convention in MadGraph. The mediator part of the
Lagrangian is thus:
\begin{equation} \label{L_R}
\mathcal{L}_R  = -\frac{1}{4} F^{\mu\nu} F_{\mu\nu} -
\frac{1}{2} m_R^2 R^\mu R_\mu\,, \ \ \ {\rm with} \ F_{\mu\nu} \equiv
\partial_\mu R_\nu - \partial_\nu R_\mu\,.
\end{equation}
In order to allow both vector and axial-vector couplings the DSP
should be a Dirac fermion, because Majorana fermions cannot have a
vector interaction. Again using MadGraph convention, the corresponding
piece of the Lagrangian is
\begin{equation} \label{L_DSP}
\mathcal{L}_{\rm DSP} = \bar{\chi}(i\slashed{\partial} - m_\chi) \chi \,.
\end{equation}
Finally, the interaction terms are
\begin{equation} \label{L_I}
  \mathcal{L}_I = \sum_{q=s,\,c}g^V_q R_\mu \bar{q}\gamma^\mu
  q+\sum_{q=b,\,t} R_\mu \bar{q} \gamma^\mu \left( g^V_q - g^A_q \gamma^5
  \right) q + R_\mu \bar{\chi} \gamma^\mu \left( g^V_\chi - g^A_\chi \gamma^5
  \right) \chi \,.
\end{equation}

In this model DSPs can scatter off nucleons via $R$ exchange. The
corresponding spin--independent and spin--dependent cross sections
are to leading order in perturbation theory \cite{Frandsen:2012rk}:
\begin{equation} \label{sig_dir}
  \sigma_{n/p}^{\rm SD} = a_{n/p}^2 \frac {3\mu^2_{n/p}} {\pi m_R^4}\,; \ \
  \sigma_{n/p}^{\rm SI} = f_{n/p}^2 \frac{3\mu^2_{n/p}}{\pi m_R^4}\,; \ \ 
  \mu_{n/p} = \frac {m_\chi m_{n/p}} {m_\chi+m_{n/p}}\,.
\end{equation}
The coefficients $a_{n/p}$ and $f_{n/p}$ depend on different combinations of
couplings:
\begin{equation} \label{fN}
f_p = g_\chi^V ( 2g_u^V + g_d^V )\,;\ \ f_n = g_\chi^V (g_u^V + 2g_d^V)\,;
\end{equation}
and \cite{Olive:2016xmw}
\begin{eqnarray} \label{aN}
  a_{n/p} &=& g^A_\chi \sum_{q=u,\,d,\,s} \Delta q^{(n/p)} g_q^A\,;\nonumber \\
  \Delta u^{(p)} &=& \Delta d^{(n)} = 0.84 \pm 0.02\,; \\
  \Delta u^{(n)} &=& \Delta d^{(p)} = -0.43 \pm 0.02\,; \nonumber\\
  \Delta s^{(p)} &=& \Delta s^{(n)} = -0.09 \pm 0.02\,. \nonumber
\end{eqnarray}
Eqs.(\ref{fN}) show that setting $g^V_{u,\,d}=0$ suffices to make the
leading order spin--independent cross sections on protons and neutrons
vanish.  On the other hand, eqs.(\ref{aN}) show that
$g^A_{u,\,d,\,s}=0$ is needed in order to ``switch off'' the leading
order spin--dependent cross sections; weak $SU(2)$ invariance then
implies $g^A_c = 0$ as well.

This leaves us with seven free parameters: $g^V_{s,\,c}$,
$g^V_{b,\,t}$, $g^A_{b,\,t}$, $g^A_\chi$, $g^V_\chi$, $m_R$ and
$m_\chi$. However, since $R$ does not couple to leptons signals
involving missing transverse energy $\slashed{E}_T$ require a pair of
DSPs in the final state. Since SM $Z$ boson couple to all quarks,
final states with an $R$ boson replaced by an invisible decaying $Z$
boson will always contribute to (and indeed often dominate) the
background to these signals. Clearly the signal can only compete with
this background from on--shell $Z$ bosons if on--shell
$R \rightarrow \chi \bar \chi$ decays are possible. The relevant
quantity is then the branching ratio for these decays, rather than the
couplings $g^V_\chi$ and $g^A_\chi$ separately. Moreover, the DSP
mass $m_\chi$ also affects the signal only through this branching
ratio. This observation implies that replacing the Dirac DSP $\chi$ by
a complex scalar $\phi$ is trivial, since again only the branching ratio
for $R \rightarrow \phi \bar \phi$ decays is relevant in that model.

Turning to quark couplings, we assume all non--vanishing couplings to
be equal. In case of axial vector couplings, this can again be
motivated by $SU(2)$ invariance. This would still allow different,
non--vanishing second and third generation vector couplings, but we
set them equal for simplicity. Note that the case $g^V_s = g^V_c = 0$
would give very similar results as the scenario with non--vanishing
axial vector couplings. The reason is that contributions to the
relevant matrix elements from $g^V_q$ and $g^A_q$ differ only by terms
of the order $m_q/Q$, where $Q$ is the energy scale of the process.
Since the parton distribution function for top quarks is still very
small at the energies we are interested in, and top tagging turns out
to be quite inefficient, the relevant quark is the $b$ quark, and
$m_b/Q \ll 1$ for all cases of interest to us. The main difference
between vector and axial vector couplings is therefore that in the former
case couplings to second generation quarks are included, while these
couplings vanish in the latter case.

Finally we are therefore left with four relevant free parameters:
$g^V_q$, $g^A_q$, ${\rm Br}(R\rightarrow\chi\bar{\chi})$ and
$m_R$. Since the parton distribution functions for second generation
quarks in the proton are significantly larger than those for third
generation (basically, $b$) quarks, for fixed size of the non--vanishing
couplings we expect much smaller total cross sections for the case
$g_q^V=0, \, g_q^A \equiv g_q$ than for the case
$g_q^A=0, \, g_q^V \equiv g_q$. On the other hand, scenarios with
$g^V=0$ should have higher efficiency for $b$ tagging, which is
required in some searches.

\subsection{Perturbativity and Unitarity Conditions}
\label{subsec:2.2}

We will use leading order tree--level diagrams in the
simulation. Therefore, the relevant couplings should not be too big, so
that the perturbation theory is reliable. We impose the simple
perturbativity condition
\begin{equation} \label{pert1}
  \Gamma_R < m_R\,,
\end{equation}
where $\Gamma_R$ is the total decay width of the mediator. The
partial width for $R \rightarrow f \bar f$ decay, where $f$ is some fermion,
is:
\begin{equation} \label{Gamma_R}
 \Gamma(R \rightarrow f\bar{f} ) = \frac {m_R N_C^f} {12\pi} \sqrt{ 1 - 4z_f }
  \left[ (g^V_f)^2 + (g^A_f)^2 + z_f \left( 2(g^V_f)^2 - 4(g^A_f)^2 \right)
   \right]\,,
\end{equation}
where $z_f\equiv m_f^2/m_R^2$ and $N_C^f = 3 \ (1)$ for
$f=q \ (f=\chi)$. Since we use ${\rm Br}(R\rightarrow\chi\bar{\chi})$
as free parameter instead of $m_\chi$ and $g_\chi^{V,A}$, the
perturbativity condition can be written as
\begin{equation} \label{pert2}
  \sum_{q} \sqrt{1-4z_q} \left[ (g^V_q)^2 + (g^A_q)^2 + z_q
    \left( 2 (g^V_q)^2 - 4 (g^A_q)^2 \right) \right] < 4\pi[1-{\rm Br}
  (R\rightarrow \chi\bar{\chi})] < 4\pi\,.
\end{equation}
In the next section, we will only discuss the mass range where the
bounds for vector or axial vector couplings are smaller than $2$; this
satisfies the perturbativity condition.

Another important theoretical constraint on the parameters in the
Lagrangian originates from demanding that unitarity in scattering amplitudes
is preserved \cite{Kahlhoefer:2015bea}. This constraints the axial
vector couplings between the mediator and fermions:
\begin{equation}
 g_f^A \frac {m_f} {m_R} \leqslant \sqrt{ \frac {\pi} {2} }\,.
\end{equation}
Due to the assumption of universal axial vector couplings to $b$ and
$t$ quarks, the strongest constraint always comes from the much
heavier top, and becomes quite strong for light mediator:
\begin{equation} \label{unitarity}
  g_q^A\leqslant\sqrt{\frac{\pi}{2}}\frac{m_R}{m_t}
  =\frac{m_R}{137.59\ {\rm GeV}}\,.
\end{equation}
For example, for $m_R=10$ GeV, $g^A$ should be smaller than $0.08$. In
contrast, for $m_R>275$ GeV the unitarity constraint becomes weaker
than the perturbativity condition.

\section{Application to LHC Data} 
\label{sec:3} 
\setcounter{footnote}{0}

In this section, we recast various LHC searches to constrain the model
introduced in section \ref{sec:2}, including a mono--jet + $\slashed{E}_T$
search \cite{Aaboud:2017phn}, multi--jet + $\slashed{E}_T$ searches
\cite{Aaboud:2017vwy, Aaboud:2017ayj, ATLAS:2016lsr}, a multi--jet +
$\slashed{E}_T$ searches with $t$ tag \cite{Aaboud:2018zpr}, a multi--jet +
$\slashed{E}_T$ search with double $b$ tag \cite{Aaboud:2017wqg}, and di--jet
resonance searches with final state $b$--jets \cite{Aaboud:2018tqo} or
$t$--jets \cite{Sirunyan:2017uhk}.

In order to simulate the events and recast the analysis, we use
FeynRules \cite{Alloul:2013bka} to encode the model and generate an
UFO file \cite{Degrande:2011ua} for the simulator, MadGraph
\cite{alwall2011madgraph} to generate the parton level events, PYTHIA
8 \cite{sjostrand2015introduction} for QCD showering and
hadronization, DELPHES \cite{deFavereau:2013fsa} to simulate the ATLAS
and CMS detectors, and CheckMATE \cite{Drees:2013wra, Dercks:2016npn}
to reconstruct and $b$--tag jets, to calculate kinematic variables, and
to apply cuts. We note that the toolkit CheckMATE uses a number of
additional tools for phenomenology research \cite{Cacciari:2011ma,
  Cacciari:2005hq, Cacciari:2008gp, Read:2002hq, Lester:1999tx,
  Barr:2003rg, Cheng:2008hk, Bai:2012gs, Tovey:2008ui,
  Polesello:2009rn, Matchev:2009ad}.

Let us first discuss final states involving missing $E_T$. These are
often categorized as ``mono--jet + $\slashed{E}_T$'' and ``multi--jet +
$\slashed{E}_T$'' final states. However, the ``mono--jet'' searches
also allow the presence of at least one additional jet. On the
other hand, ``multi--jet'' searches do indeed require at least two
jets in the final state. These signals thus overlap, but are not identical
to each other.

\begin{figure}[thb]
  \includegraphics[width=\textwidth]{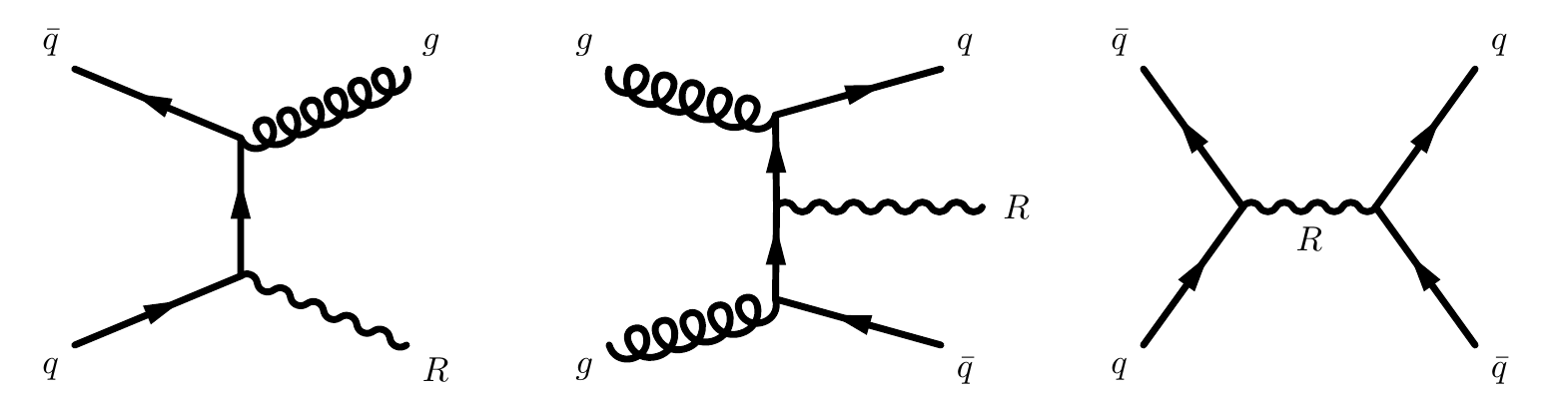}
  \caption{Examples of Feynman diagrams contributing to mono--jet +
  $\slashed{E}_T$ (left), di--jet + $\slashed{E}_T$ (center) and
    di--jet resonance (right) final states; in the former two cases it
    is assumed that the mediator $R$ decays into two dark sector
    particles, which escape detection, whereas in the latter case $R$
    is assumed to decay into a quark antiquark pair. The diagram to
    the right is unique (with different initial states contributing),
    and the one on the left is unique up to crossing; however, many
    additional diagrams, with different combinations of partons in the
    initial and final states and different propagators, contribute
    to $R + $di--jet production.}
\label{fig:feyn}
\end{figure}

As remarked in the Introduction, missing $E_T$ in signal events always
comes from invisibly decaying mediators,
$R \rightarrow \chi \bar \chi$. Since multi--jet searches require at
least two jets in the final state, we use MadGraph to generate
parton--level events with a $\chi \bar \chi$ pair plus one or two
partons (quarks or gluons) in the final state. The former process only
gets contributions from the left diagram in fig.~\ref{fig:feyn} plus
its crossed versions, including the contribution from
$g q \rightarrow R q$. Note that $R$ has to couple to the initial
quark line in this case. We use parton distribution functions (PDFs)
with five massless flavors; the mass of the corresponding quarks
should be set to $0$ in order to avoid the inconsistency with massless
evolution equations (DGLAP equations). The $b-$quark PDF is nonzero,
but it is still considerably smaller than those of first generation
quarks. The contribution from this diagram, which is formally of
leading order in $\alpha_S$, is therefore quite small, especially for
scenarios with $g^V_q = 0$ where $R$ only couples to third generation
quarks.

If we allow the final state to contain two partons in addition to the
DSPs, there are contributions with only light quarks or gluons in the
initial state; an example is shown in the middle of
fig.~\ref{fig:feyn}, but there are several others. These diagrams are
higher order in $\alpha_S$, but they receive contributions from
initial states with much larger PDFs than those contributing to the
first diagram. It is thus not clear a priori which of these
contributions will be dominant for a given set of cuts.

There is one additional complication. At the parton level, events with
one and two partons in the final state are clearly distinct. However,
once we include QCD showering, which is handled automatically by
PYTHIA, the distinction becomes less clear. In particular, a single
parton event with an additional gluon from showering can no longer be
distinguished from a certain two parton event without additional
gluon. Naively adding contributions with one and two partons in the
final state before showering can therefore lead to double
counting. Similarly, if one of the final--state quarks shown in the
middle diagram of fig.~\ref{fig:feyn} has small $p_T$, the diagram can
be approximated by $g \rightarrow q \bar q$ splitting followed by
$gq \rightarrow Rq$ production. This contribution is already contained
in the crossed version of the left diagram of fig.~\ref{fig:feyn}, via
the scale--dependent PDF of $q$, so simply adding these diagrams again
leads to double counting.  MadGraph avoids both kinds of double
counting by using the ``MLM matching'' algorithm
\cite{Mangano:2006rw}. Of course, showering can add more than one
additional parton; indeed, we find significant rates for final states
with up to four jets (having transverse energy $E_T \geq 35$ GeV
each).

Searches for final states leading to large missing $E_T$ are typical
cut--and--count analyses, where the final state is defined by cuts on
the type and number of final state objects (in particular, leptons and
jets with or without $b-$tag) and on kinematic quantities (in
particular, the transverse momenta or energies of the jets and the
missing $E_T$). The experiments themselves designed these cuts, and
estimated the expected number of surviving SM background events. The
comparison with the actually observed number of events after cuts then
allows to derive upper bounds on the number of possible signal events.
We pass our simulated signal events through CheckMATE, which applies
the same cuts (including detector resolution effects), and compares
the results with the upper bounds obtained by the experiments.

The second kind of search we consider are searches for di--jet
resonances. The leading--order signal diagram is shown on the right in
fig.~\ref{fig:feyn}. In this case the final state contains no partons
besides the mediator $R$; for $g^V_q = 0$, only $b \bar b$ initial
states contribute, whereas for non--vanishing vector couplings also
$s \bar s$ and $c \bar c$ initial states contribute. Of course, the
left and middle diagrams shown in fig.~\ref{fig:feyn} also contribute
to this signal if $R$ decays into a $q \bar q$ pair. However, in this
case one has to add two powers of $\alpha_S$ in order to access
initial states including only light quarks or gluons. Moreover, if all
final state transverse momenta are small, which maximizes the cross
section, the contribution from the middle diagram is actually already
included in the right diagram, via double $g \rightarrow q \bar q$
splitting. The left and middle diagrams should therefore only be
included in inclusive $R$ production when a full NLO or even NNLO
calculation is performed, which is beyond the scope of this work.

Note also that resonance searches are not cut--and--count
analyses. The analyses still use a set of basic acceptance cuts, in
this case on the (pseudo--)rapidities and transverse momenta of the
two leading jets. The bound on resonance production is then obtained
by fitting a smooth function to the di--jet invariance mass
distribution, which is assumed to be dominated by backgrounds, and
computing the limit on a possible additional contribution peaked at a
certain value (basically, the mass of the resonance). The current
version of CheckMATE does not include comparison with this kind of
searches.  However, CheckMATE does allow to estimate the efficiency
with which our signal events pass the acceptance cuts. This allows to
derive the constraints from resonance searches on our model, as
follows.

The most sensitive di--jet resonance search we found is that of
ref.~\cite{Aaboud:2018tqo}, which requires a double $b-$tag in the
final state. This paper presents the resulting upper bounds for a
couple of models.  One of them is quite similar to ours, but assumes
universal couplings to all quarks; this leads to a greatly enhanced
resonance production cross section, and a somewhat reduced branching
ratio into $b \bar b$ pairs, compared to our model. The paper also
gives the cut efficiency for the model with universal couplings. We,
therefore, recast their cuts and compare the cut efficiencies of their
model and our models in order to estimate the bound for our model
through the following rescaling:
\begin{equation} \label{rescale}
  \sigma_{\rm max, \, ours} = \sigma_{\rm max, \, exp}
  \cdot \frac{\epsilon_{\rm exp}} {\epsilon_{\rm ours}}\,.
\end{equation}
Here $\sigma_{\rm max, \, ours}$ is the largest allowed cross section
for our model, $\sigma_{\rm max, \, exp}$ is the largest allowed cross
section in the original experimental analysis, $\epsilon_{\rm exp}$ is
the selection efficiency of the model in the paper, and
$\epsilon_{\rm ours}$ is the selection efficiency of our
model.

Finally, we cannot easily reproduce the top tagging required in the
di--top resonance search~\cite{Sirunyan:2017uhk}. However, even if we
assume $100\%$ efficiency for the di--top tag, the resulting
bound is much weaker than our recast of \cite{Aaboud:2018tqo} described in
the previous paragraphs. We therefore do not show this bound in our
summary plot.

\begin{figure}[htb]
\includegraphics[width=0.5\textwidth]{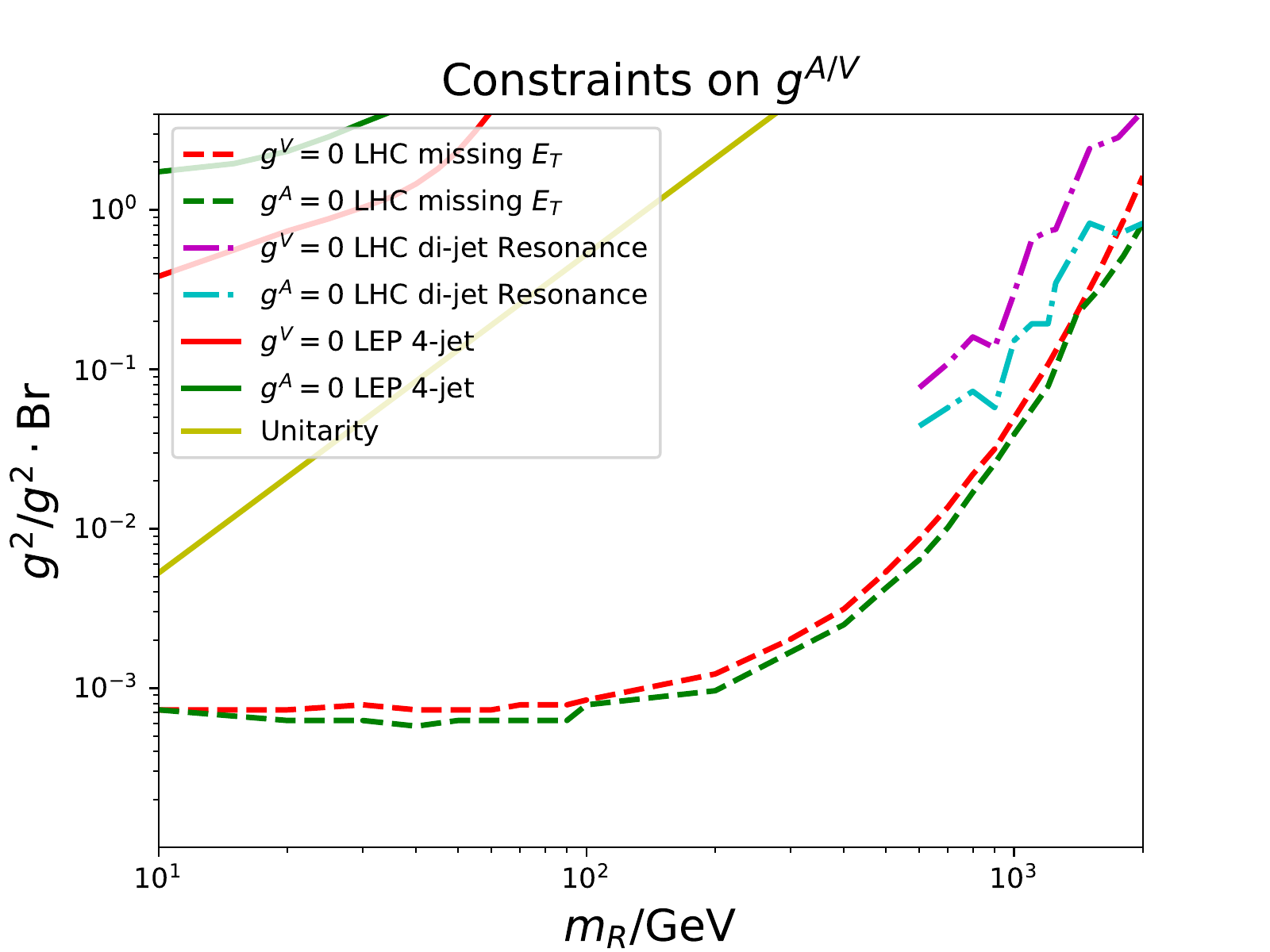}
\includegraphics[width=0.5\textwidth]{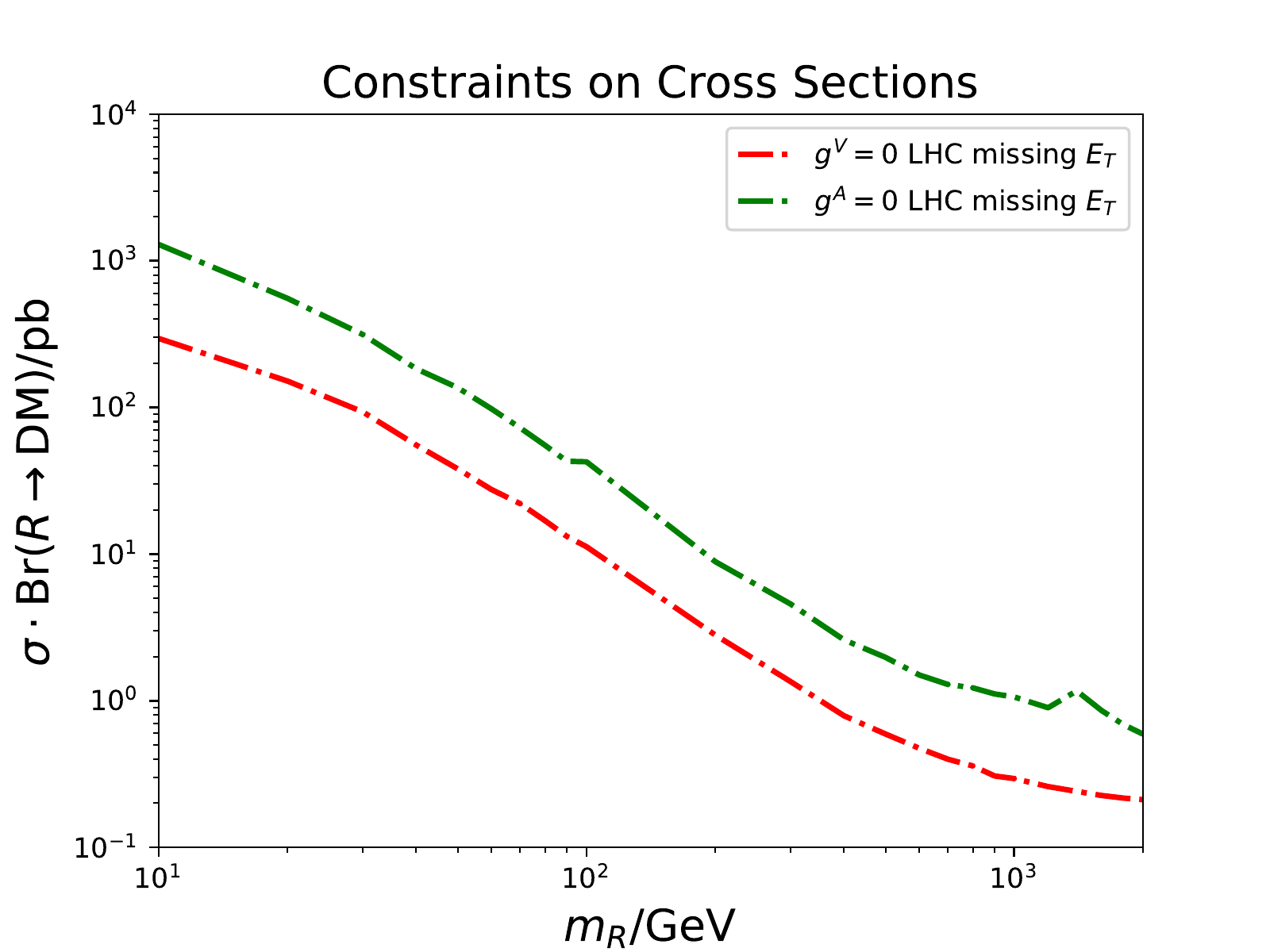}\\
\caption{The left frame shows the bound on the squared coupling of the
  mediator to quarks times the relevant branching ratio of the
  mediator that we derive by recasting various LHC bounds (dashed and
  dot--dashed curves). The lower solid curve shows the unitarity bound
  (\ref{unitarity}) on the axial vector coupling, and the thin solid
  lines in the top--left corner show constraints from recast LEP
  constraints \cite{Drees:2018gvr} based on analyses by the ALEPH
  collaboration. For $g^A=0$ only vector couplings $g^V_q$ are allowed
  with $q = s, c, b, t$, while for $g^V=0$ only axial vector couplings
  $g_q^A$ are allowed with $q = b, t$. LHC missing $E_T$ results are
  from the combination of mono--jet and multi--jet analyses. The right
  frame shows the upper bound on the total cross sections from the
  missing $E_T$ analyses.}
\label{rst:01}
\end{figure}

The results of our analyses are summarized in fig.~\ref{rst:01}. The
thin solid lines in the top--left corner show the bounds we derived
\cite{Drees:2018gvr} from analyses of older ALEPH searches for four
jet final states at the $e^+e^-$ collider LEP; note that these bounds
are valid for $m_R < 2 m_\chi$. The solid straight line is the
unitarity bound (\ref{unitarity}) applied to the top mass; recall that
it applies only to axial vector couplings. (Since top quarks could not
be produced at the LEP collider, in \cite{Drees:2018gvr} we only
considered the unitarity constraints involving $m_b$ and $m_\chi$.)

The other results shown in fig.~\ref{rst:01} are new. The dashed
curves show the bounds on the square of the coupling of $R$ to quarks
times the branching ratio for invisible $R$ decays which we derived
from the most sensitive jet(s) plus missing $E_T$ searches, for pure
axial vector couplings (red, upper curve) and pure vector couplings
(green, lower curve); the right frame shows the corresponding bounds
on the signal cross section, defined as the total cross section for
the on--shell production of a mediator $R$ times the invisible
branching ratio of $R$. It is important to note that these constraints
are only significant in our model if on--shell
$R \rightarrow \chi \bar \chi$ decays are allowed, i.e. they constrain
a region of parameter space that is complementary to that analyzed in
ref.~\cite{Drees:2018gvr}.

The dot--dashed curves in the left frame show the bounds on the square
of the coupling of $R$ to quarks times the branching ratio for
$R \rightarrow q \bar q$ decays that result from searches for di--jet
resonances, again separately for pure axial vector couplings (purple,
upper curve) and pure vector couplings (blue, lower curve). The
relevant analysis by the ATLAS collaboration \cite{Aaboud:2018tqo} is
sensitive only to $m_R \geq 600$ GeV.

The difference between the constraints on vector and axial vector couplings
is almost entirely due to the additional coupling to $s$ and $c$ quarks that
we allow only for the former, as discussed in Sec.~2.1. In particular, we see
that the constraint from the $b \bar b$ resonance search is much stronger for
the model with vector couplings.

In the left frame of Fig.~\ref{rst:01} the curves depicting the bounds from
searches for final states containing $\slashed{E}_T$ evidently lie below the
ones showing bounds from di--jet resonance searches, except for the scenario
with pure vector coupling at $m_R \simeq 2$ TeV. However, this is somewhat
misleading, since the dashed curves show bounds on $g_q^2 \cdot {\rm
Br}(R\rightarrow\chi\bar{\chi})$, while the dot--dashed curves shows bounds
on $g_q^2 \cdot [ 1 - Br(R\rightarrow\chi\bar{\chi})]$. For $m_R \geq 1$ TeV
the two sets of constraints on the coupling are actually comparable if ${\rm
Br}(R\rightarrow\chi\bar{\chi}) \simeq 0.3 \ (0.1)$, for pure vector (axial
vector) coupling; for even smaller invisible branching ratio of $R$, the $b
\bar b$ resonance search imposes the stronger constraint in this large $m_R$
region. We note that for $m_R^2 \gg m_t^2$ and $g_\chi = g_q$, i.e. equal
coupling of the mediator to the DSP and to heavy quarks, the invisible
branching ratio of $R$ is below $1/7 \ (1/13)$ for pure axial vector (vector)
coupling, the difference being due to the different number of accessible $q
\bar q$ final states.

Within the missing $E_T$ searches the best bound on $g^V_q$ for $m_R<1.4$ TeV
is from ref.\cite{Aaboud:2017wqg}, a double $b$ tagged multi--jet +
$\slashed{E}_T$ analysis, while ref.\cite{Aaboud:2017vwy}, a general
multi--jet + $\slashed{E}_T$ analysis, is the most sensitive one for $m_R\geq
1.4$ TeV; this change of the most sensitive analysis explains the structure
in the dark green curves at that $m_R$, which is most visible in the right
frame. In contrast, the strongest bound on $g^A_q$ is always from
ref.\cite{Aaboud:2017wqg} with double $b$ tag, which also determines the
bound on the vector coupling for $m_R < 1.4$ TeV. This explains why the bound
on the coupling is actually very similar in both cases: the required double
$b$ tag means that the contribution from partonic events containing only $s$
or $c$ quarks, which only exists in the case of vector coupling, has very
small efficiency, since the $b$ tag requirement can only be satisfied though
mistagging, or through additional $b$ quarks produced in hard showering. As a
result the bound on the {\em total} cross section, shown in the right frame,
is much weaker for pure vector coupling, since the coupling to $s$ and $c$
quarks greatly increases the total cross section while contributing little to
the most sensitive signal.

We also consider multi--jet analyses specially designed for final
states containing two top quarks \cite{Aaboud:2017ayj,
  Aaboud:2018zpr}. However, the top--tag in \cite{Aaboud:2018zpr} is
not easy to recast directly. We therefore, assume $100\%$ efficiency
to reach the most ideal bound. Unfortunately, even this ideal bound on
$g^2$ is $10$ times weaker than that from the analysis which only
requires a double $b$-tag. One reason is that both $b \bar b$ and
$t \bar t$ final states may lead to $b$-tagged jets, while the
selection rules specially designed for top jets exclude the $b\bar b$
final state. Moreover, for our assumption of equal couplings the cross
section for $t \bar t R$ production is considerably smaller than that
for $b \bar b R$ production.

As noted above, we also derived constraints on our model from mono--jet
searches. The most sensitive analysis has been published
in~\cite{Aaboud:2017phn}, and does not require any flavor
tagging. The resulting constraint on the vector coupling is only
slightly weaker than that shown in Fig.~\ref{rst:01}, while the
constraint on the axial vector coupling is not competitive. Since no
flavor tagging is required, the large contribution from $s$ or $c$
quarks in the initial and final states has similar efficiency as
contributions with $b$ quarks, and greatly strengthens the limit on
the vector coupling.

\begin{figure}[thb]
  \includegraphics[width=\textwidth]{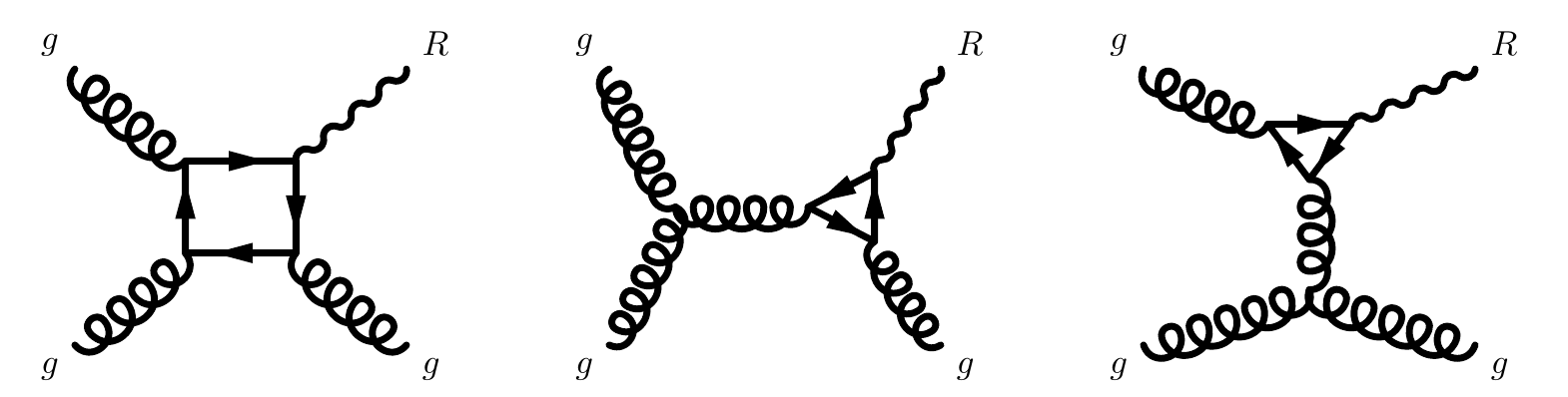}
  \caption{Loop diagrams contributing to $R+$jet production from gluon
    fusion.}
\label{fig:feynLoop}
\end{figure}

Before concluding this section, we comment on loop processes that
allow $gg$ initial states to contribute to our signals. The relevant
Feynman diagrams are shown in Fig.~\ref{fig:feynLoop}. They involve
two additional QCD vertices relative to the leading--order $R+$jet
production channels, i.e. they are formally NNLO. Nevertheless the
large gluon flux in the proton might lead to sizable contributions.
We again use FeynRules and Madgraph to simulate these events at the
parton level.

Note that the tree--level contributions we discussed so far are only
sensitive to the absolute value of the couplings of the mediator to quarks.
In contrast, in the loop diagrams all quark flavors contribute
coherently, so the relative signs between different $R q \bar q$ couplings
are important.

Let us first consider pure vector couplings. Here our simplified model
as written is well--behaved also at QCD one--loop level. We find that
the loop contributions of Fig.~\ref{fig:feynLoop} only contribute
at most $2\%$ of the leading--order mono--jet signal if all $g^V_q$ are set
equal; this contribution is reduced by another factor of $5$ if we
instead take $g^V_{s/b}=-g^V_{c/t}$. In particular, there is no
enhancement for small $m_R$; instead, the cross section after cuts
approaches a constant once $m_R \ll p_{T,{\rm cut}}$. Recall also that
in this case there are tree--level contributions involving the strange
quark content of the proton, which is much larger than that of $b$
quarks (although still considerably smaller than that of gluons). We
can thus always safely neglect these loop contributions for non--zero
vector couplings.

In contrast, in case of non--vanishing axial vector couplings our
model with equal couplings of the mediator to all heavy quarks leads
to a $ggR$ anomaly, i.e. this version of our simplified model is not
well--behaved at the 1--loop level. We therefore took $g^A_b=-g^A_t$
in order to cancel this anomaly.

Moreover, the loop amplitude now receives a contribution that scales
$\propto 1/m_R$. As a result, for $m_R = 10$ GeV the loop contribution
to the mono--jet cross section exceeds the tree--level contribution by
about a factor of $20$. We find that nevertheless the best bound still
comes from the final state with two $b-$jets and missing $E_T$. Recall
that here $gg$ initial states are accessible already at
tree--level. Since in the loop diagrams the external gluon has to be
virtual, so that it can split into a $b \bar b$ pair, the loop
contribution is still NNLO relative to this tree--level
contribution. Nevertheless the $1/m_R$ enhancement, which is
associated with heavy (i.e. top) quark loops, means that including the
loop diagram with $g^A_b=-g^A_t$ would tighten the upper limit on the
squared coupling shown in Fig.~\ref{rst:01} at $m_R=10$ by about
$40\%$. For $m_R=50$ GeV, however, the loop contribution only doubles
the total mono--jet signal, and the final bound on the squared coupling from
the di-$b$ final state is improved by about $3\%$.

It should be clear that setting $g^A_b=-g^A_t$ is only one solution to
cancel the anomaly. Another possibility is to introduce a very heavy
quark $Q$ satisfying $g^A_Q=-2g^A_b=-2g^A_t$. This would lead to even
larger loop contributions for small $m_R$; however, the unitarity
bound (\ref{unitarity}) would then also have to be applied to $m_Q$,
and might even supersede the LHC constraint.

In sum, we conclude that for axial vector couplings loop corrections
involving two--gluon initial states might moderately strengthen the
LHC constraint for $m_R<50$ GeV, the exact result depending on the UV
completion of the model. Note also that this source of loop
corrections adds incoherently to the signal, i.e. it cannot weaken the
bounds presented in Fig.~\ref{rst:01}. We therefore believe that this
Figure, which is independent of the UV completion, is a better
representation of the LHC constraints on our model.

\section{Conclusions}
\label{sec:4}

In this study, we discuss a model containing a Dirac fermion $\chi$ as
dark matter candidate as well as a spin$-1$ mediator $R$. We assume that
$R$ has vanishing couplings to first generation quarks and vanishing
axial vector coupling to second generation quarks, thereby easily
satisfying constraints from direct dark matter searches. By assuming
vanishing couplings to leptons the otherwise most sensitive LHC
searches, based on analyses of $\ell^+\ell^-$ final states where
$\ell$ stands for a charged lepton, are evaded as well. Due to the
vanishing couplings to light quarks, and hence to all valence quarks
in the proton, the $R$ production rate at the LHC is considerably
smaller than for the more commonly considered scenarios with
(essentially) universal couplings to all quarks.

Nevertheless LHC data impose quite strong constraints on the model if
the branching ratio for invisible $R$ decays is sizable, which
requires $m_R > 2 m_\chi$. The best LHC bound then always comes from
searches for final states containing jets plus missing $E_T$. Our
CheckMATE--based recast of these analyses leads to an upper bound on
the product of the squared coupling and the invisible branching ratio
of $R$ of $10^{-3}$ for $m_R \leq 200$ GeV. This weakens to $0.01$
($1$) for $m_R = 600$ GeV ($2$ TeV), see Fig.~\ref{rst:01}. Searches
for invisibly decaying mediators have traditionally been framed as
``mono--jet'' searches (which allow additional jets in the final
state, as mentioned above), and have been interpreted assuming equal
(vector or axial vector) couplings to all quarks
\cite{Sirunyan:2017hci, Aaboud:2017phn}. For pure axial vector
couplings these bounds are actually weaker than ours if
$m_R \leq 600$ GeV. Since the signal need only contain a single hard
jet, and no $b-$tagging is used, one needs a very strong cut on the
missing $E_T$ to suppress the background; for $m_R \lsim 1$ TeV this
leads to a much worse cut efficiency than the most sensitive analysis
we use, which requires two tagged $b-$jets plus missing $E_T$. For
$m_R \lsim 600$ GeV this search may thus also impose tighter bounds on
the model with universal couplings. Nevertheless the bound on $g_q^2$
times the invisible branching ratio from mono--jet searches in the model
with universal coupling becomes significantly stronger than ours for
larger $m_R$, by about one order of magnitude for $m_R = 1.5$ TeV.

For $m_R \geq 0.6$ TeV roughly comparable bounds on the product of the
squared coupling and the branching ratio of $R$ into $q \bar q$ quarks
can be derived from an ATLAS search for $b \bar b$
resonances. Searches for generic di--jet or $t \bar t$ resonances
yield much weaker constraints on our model. Generic di--jet resonance
searches at the 13 TeV LHC become sensitive only at a resonance mass
above 1.5 TeV or so. The resulting bounds on mediators with
unsuppressed couplings to valence quarks are quite strong. For
example, for $m_R = 1.5$ TeV the ATLAS analysis \cite{Aaboud:2017yvp}
gives a bound on the squared universal coupling to quarks in a
leptophobic model that is about two orders of magnitude stronger than
our bound from $b \bar b$ resonant searches in the model with vector
couplings, which in turn is a factor of about $3$ stronger than the
analogous bound in the model with axial vector couplings.

We thus see that both in the missing $E_T$ and in the resonance
searches switching off the couplings to first generation quarks
greatly weakens the limits on the couplings for $m_R > 1$ TeV, less so
for smaller mediator masses.

Since the energy scale of these reactions (e.g. the missing $E_T$, or
$m_R$ in the resonance searches) is much larger than the masses of the
relevant quarks, the matrix elements for vector and axial vector
couplings are almost the same.  Unless $m_R \gg m_\chi$ for equal
coupling strengths the branching ratio for invisible
$R \rightarrow \chi \bar\chi$ decays will be larger for pure vector
coupling than for pure axial vector coupling; however, this effect is
absorbed by interpreting the relevant constraints as upper bounds on
the product of the squared coupling times the invisible branching
ratio, as we did in the above discussion.

LHC searches lose sensitivity to our model if $m_R > 2$ TeV, or if
$m_R < 0.6$ TeV and $m_R < 2 m_\chi$. Probing significantly higher
values of $m_R$ would require higher center--of--mass energies; since
all relevant searches are background--limited, increasing the
luminosity will increase the reach only slowly. If on--shell
$R \rightarrow \chi \bar \chi$ decays are not possible, missing $E_T$
searches at the LHC are essentially hopeless in our model. The reason
is that in this case a signal which is of second order in the
couplings of the mediator has to compete with SM signals that are
first order in electroweak couplings, in particular the production of
$Z$ and $W$ bosons which decay into neutrinos.\footnote{In case of
  universal couplings to all quarks the ``mono--jet'' analyses
  \cite{Aaboud:2017phn, Sirunyan:2017hci} do exclude a small region of
  parameter space with $m_R/2 < m_\chi \lsim 200$ GeV for a vector
  mediator, but not for an axial vector mediator.}  For $m_R < 70$ GeV
the old LEP experiments have some sensitivity, but the resulting bound
is not very strong \cite{Drees:2018gvr}. Straightforward di--jet
resonance searches at the LHC are not possible for $m_R$ much below
$0.6$ TeV, since the trigger rate would be too high. One might
consider previous hadron colliders, in particular the
Tevatron. However, these earlier colliders were $p \bar p$ colliders,
where the $b \bar b$ background includes contributions where both
initial--state quarks are valence quarks; recall that in our model the
signal does not receive contributions from such initial states.

A probably more promising approach is to consider final states
containing an additional hard ``tagging jet'' besides the mediator
$R$. Both ATLAS \cite{Aaboud:2018zba} and CMS \cite{Sirunyan:2017nvi}
have presented bounds on rather light di--jet resonances using this
trick, which is also employed in the ``mono--jet'' searches.
Unfortunately these searches are currently not easy to recast, since
they use ``fat jet'' substructure techniques. In any case, in order to
gain sensitivity to our model this technique would probably have to be
combined with $b-$tagging, which proved crucial for deriving useful
constraints from di--jet resonance searches at $m_R > 600$ GeV. An
analysis of this kind should be able to probe deep into the parameter
space with $m_R < 600$ GeV and $m_R < 2 m_\chi$.

\FloatBarrier

\Acknowledgements 

This work was partially supported by the by the German ministry for
scientific research (BMBF).

\bibliographystyle{unsrt}

\end{document}